\documentclass[aps, prd, 10pt, twocolumn, superscriptaddress,noshowpacs, preprintnumbers,longbibliography,nofootinbib,nobibnotes,floatfix]{revtex4-2}

\usepackage{multirow}
\usepackage{graphicx}
\usepackage{dcolumn}
\usepackage{footmisc}
\usepackage{amsmath}
\usepackage{slashed}
\usepackage{physics}
\usepackage{amssymb}

\usepackage{bm}
\usepackage{hyperref}
\usepackage{xcolor}
\usepackage{float}
\usepackage[normalem]{ulem}
\usepackage{multirow}

\begin{document}
\setlength{\abovedisplayskip}{5pt}
\setlength{\belowdisplayskip}{5pt}
\setlength{\abovedisplayshortskip}{5pt}
\setlength{\belowdisplayshortskip}{5pt}

\def\hc{\text{h.c.}}

\newcommand{\jtcc}[1]{\textcolor{violet}{{\bf JTC}: #1}}
\newcommand{\jtc}[1]{\textcolor{violet}{\bf #1}}
\newcommand{\mc}[1]{\textcolor{orange}{{\bf MC}: #1}}
\newcommand{\gh}[1]{\textcolor{blue}{{\bf GH}: #1}}
\newcommand{\tb}[1]{\textcolor{teal}{{\bf TB}: #1}}
\newcommand{\tbc}[1]{\textcolor{teal}{#1}}
\newcommand{\tbcite}{\textcolor{teal}{[cite]}}

\newcommand{\pmmc}[1]{{\color{orange} \bf [PMM: #1]}}
\newcommand{\pmm}[1]{{\color{orange} \bf #1}}

\newcommand{\greekbf}[1]{\boldsymbol{\mathbf{#1}}}
\newcommand{\mchi}{m_\chi}
\newcommand{\mphi}{m_\phi}
\newcommand{\Enuin}{E_{\nu_{\rm in}}}
\newcommand{\Enuout}{E_{\nu_{\rm out}}}
\newcommand{\mzp}{m_{Z^\prime}}

\preprint{}

\title{The Highest-Energy Neutrino Event Constrains Dark Matter-Neutrino Interactions}

\author{Toni Bertólez-Martínez}
\email{antoni.bertolez@fqa.ub.edu}
\affiliation{Departament de F\'isica Qu\`antica i Astrof\'isica and Institut de Ci\`encies del Cosmos, Universitat de Barcelona, Diagonal 647, E-08028 Barcelona, Spain}

\author{Gonzalo Herrera}
\email{gonzaloherrera@vt.edu}
\affiliation{Center for Neutrino Physics, Department of Physics, Virginia Tech, Blacksburg, VA 24061, USA}

\author{Pablo Martínez-Miravé}
\email{pablo.mirave@nbi.ku.dk}
\affiliation{Niels Bohr International Academy and DARK, Niels Bohr Institute, University of Copenhagen,\\
Blegdamsvej 17, 2100, Copenhagen, Denmark}

\author{Jorge Terol Calvo}
\email{jortecal@protonmail.com}
\affiliation{Istituto Nazionale di Fisica Nucleare, Sezione di Torino, I-10125, Torino, Italy}

\begin{abstract}
Dark Matter-neutrino interactions affect the propagation of astrophysical neutrinos, attenuating the flux of neutrinos arriving at Earth. Using the highest-energy neutrino event detected to date by the KM3NeT collaboration as an example, and assuming an extragalactic origin, 
we derive limits on these interactions at $E_\nu = 220^{+570}_{-110}\, \mathrm{PeV}$. Considering only the propagation on the Milky Way Dark Matter halo, we constrain the interaction cross section over the mass of the Dark Matter candidate to be, $\sigma_{\rm DM-\nu}/m_{\rm DM} \lesssim 10^{-22}\, \mathrm{cm}^2\,\mathrm{GeV}^{-1}$. If a transient source was positively identified, the high-energy neutrino would have crossed the dark-matter halo of the source host as well, resulting in more stringent constraints (e.g., up to $ \sim$ 6 orders of magnitude assuming the blazar PKS 0605-085 is the source). These bounds on the Dark Matter-neutrino interaction cross section are translated into limits on the mass of the Dark Matter candidate, the mass of the mediator, and the coupling strength for different simplified models. We find that the constraints from the KM3-230213A high-energy event for these simplified models are almost entirely ruled out for masses above the MeV by unitarity constraints. Therefore, the attenuation of such energetic neutrinos by Dark Matter calls for richer dark sectors in order to produce meaningful constraints.
\end{abstract}

\maketitle

\section{Introduction}

In the quest to understand the nature of Dark Matter (DM), a wide range of candidates have been proposed, spanning several orders of magnitude in mass ~\cite{Hui:2016ltb, Bertone:2018krk, Green:2020jor, Cirelli:2024ssz}.
The combination of direct detection and indirect detection searches puts pressure on traditional electroweakly interacting DM candidates coupled to charged leptons and quarks \cite{LZ:2022lsv, Essig:2022dfa, Slatyer:2021qgc}. However, these constraints are significantly relaxed in models where the DM only couples to neutrinos at tree level, due to loop-suppressed couplings to charged leptons and quarks. Since the existence of DM and neutrino non-zero masses provide two strong pieces of evidence of physics beyond the Standard Model, it is important to explore potential connections --for instance, neutrino mass models featuring a stable DM candidate~\cite{Tao:1996vb,Ma:2006km}-- which generically predict sizable interactions among neutrinos and DM that can be connected to the relic abundance of DM in the Universe \cite{Lindner:2010rr,Cherry:2014xra,Berlin:2018ztp,Arguelles:2019ouk, Blennow:2019fhy}.

DM-neutrino interactions predict rich phenomenology in astrophysical and cosmological probes. For instance, these
interactions could have implications on structure formation~\cite{Boehm:2000gq, Boehm:2001hm, Mangano:2006mp, Escudero:2015yka, Olivares-DelCampo:2017feq, Akita:2023yga, Crumrine:2024sdn}. First, DM can scatter off neutrinos and suppress the matter power spectrum, similarly to warm DM. In addition, interactions with DM reduce the free-streaming length of neutrinos and thus enhance the formation of large-scale structures. Furthermore, such interactions could help ameliorate the tension in some cosmological observables~\cite{DiValentino:2017oaw,Hooper:2021rjc,Brax:2023tvn, Zu:2025lrk}.

Conversely, astrophysical observations can also be used to constrain DM-neutrino interactions. For instance, DM may annihilate or decay into neutrinos in the Galactic halo or extragalactic sources \cite{Lindner:2010rr,Arguelles:2019ouk, Suliga:2024hgp}, or DM could mediate neutrino self-interactions~\cite{Das:2022xsz}. Another scenario consists in the DM of the Universe \textit{scattering off} astrophysical neutrinos, which may induce an attenuation and a time-delay in the observed neutrino fluxes. This scenario can provide leading constraints in models where the DM-neutrino scattering cross section increases with the incoming energy of the neutrino \cite{Arguelles:2017atb,Choi:2019ixb, Ferrer:2022kei, Cline:2022qld, Fujiwara:2023lsv, Cline:2023tkp,Fujiwara:2024qos, Zapata:2025huq}.

The recent detection of an exceptionally high-energy event by the KM3NeT Collaboration~\cite{KM3NeT:2025npi}, designated KM3-230213A, has drawn the attention of the astrophysics and particle physics communities~\cite{Li:2025tqf,Fang:2025nzg,Satunin:2025uui,Muzio:2025gbr,KM3NeT:2025mfl,Cattaneo:2025uxk,Boccia:2025hpm,Borah:2025igh,Brdar:2025azm,Kohri:2025bsn,Narita:2025udw,Filipovic:2025ulm,Wang:2025lgn,He:2025bex,Neronov:2025jfj,Airoldi:2025opo,Dev:2025czz,Farzan:2025ydi}. This event was observed by the ARCA detector on February 13, 2023, coming from $l = 216.06^\circ$,  $b = -11.13^\circ$, in galactic coordinates, with angular uncertainty R(68\%)$\ = 1.5^\circ$. The event is consistent with an incoming muon neutrino with energy $E_\nu = 220^{+570}_{-110}\ \rm PeV$, representing the highest neutrino energy detected to date.

In this \textit{Article}, we leverage the observation of KM3-230213A to place constraints on DM-neutrino interactions, considering that the event corresponds to a neutrino with extragalactic origin.  We translate the bounds on the attenuation of the flux into limits on the coupling strength for different simplified models for motivated ranges of the DM and mediator masses, and find that in the energy and flux of this event such simplification leads to bounds which are mostly ruled out by unitarity. 
For completeness, in the appendices we present a detailed calculation of the relevant DM column density and the DM-neutrino interaction cross sections for several models of interest for this work.

\section{Neutrino attenuation from DM interactions}
\label{sec:attenuation}

The scattering of astrophysical neutrinos with particles along their path to Earth can affect the observed flux. While neutrinos can scatter off photons, protons, and other Standard Model particles; in order to place limits we assume that DM-neutrino scattering dominates.

In this scenario, the evolution of the flux along the path can be computed by solving the cascade equation~\cite{Vincent:2017svp,Arguelles:2017atb,Cline:2022qld}
\begin{equation}
    m_{\rm DM}\dfrac{\dd{\Phi} (E)}{\dd{\Sigma_{\mathrm{DM}}}} = -\sigma(E) \Phi(E) + \int^\infty_E \dd{\Tilde{E}}\, \dfrac{\dd{\sigma(\Tilde{E},E)}}{\dd{E}}\, \Phi(\Tilde{E}), \label{eq:Cascade}
\end{equation}

where $E$ is the neutrino energy, $\sigma(E)$ is the neutrino-DM total cross section for a neutrino with energy $E$, $\dd{\sigma(\Tilde{E},E)}/\dd{E} $ is the differential cross section between incoming energy $\Tilde{E}$ and outgoing energy $E$, and $\Sigma_{\rm DM}$ is the DM column density, see the Supplementary Material for details.
%
%

The first term in the RHS of Eq.~\eqref{eq:Cascade} accounts for the loss of neutrinos off the flux at a given neutrino energy, while the second accounts for the redistribution of higher-energy neutrinos down-scattered to those energies, changing the spectrum's shape. 
The second term in the cascade equation is subdominant in models where the DM-neutrino scattering cross section rises with energy $\sigma \sim E^{n}$ less steeply than the neutrino flux decreases with energy $\Phi \sim E^{-\alpha}$, i.e. for $n<\alpha$.
Since most DM models predict $n \leq 2$, while high-energy neutrino fluxes should fall as $\alpha \simeq 2-3$~\cite{Fang:2017zjf}, we can neglect the second term in Eq.~\eqref{eq:Cascade}.

Then, the flux of neutrinos produced at a source is attenuated at a given neutrino energy $E_{\nu}$ by interactions with DM on their way to the Earth as
\begin{align}
    \frac{\Phi^{\rm obs}_{\nu}}{\Phi^{\rm em}_\nu}= e^{-\sigma\cdot \frac{\Sigma_{\rm DM}}{m_{\rm DM}}}
    \, .
\end{align}
Here, $\Phi^{\rm obs}_i$ and $\Phi^{\rm em}_i$ are respectively the observed and emitted fluxes of neutrinos,
and $\sigma\cdot \frac{\Sigma_{\rm DM}}{m_{\rm DM}}$ is an attenuation coefficient from DM-neutrino interactions, which is in general dependent on $E_{\nu}$.

Extragalactic neutrinos will cross different DM regions on their way to Earth, namely the DM halo in the host galaxy, the intergalactic medium and the Milky Way's halo, i.e.,
\begin{align}
\Sigma_{\rm DM} =
\Sigma_{\rm DM}^{(\mathrm{host})}+
\Sigma_{\rm DM}^{(\mathrm{intergalactic})}+
\Sigma_{\rm DM}^{(\mathrm{MW})}\, .
\end{align}
	
with typical values for the MW halo ranging between $\sim10^{22}$ and $\sim10^{23}$ GeV/cm$^2$, and the precise value in the direction of KM3-230213A will be computed in the next section. 
Upper bounds on the cross section can be obtained by imposing that the attenuation by DM-neutrino interactions is less than 90\%~\cite{Ferrer:2022kei, Choi:2019ixb, Kelly:2018tyg, Alvey_2019} at a given $E_{\nu}$. This translates into an upper limit on the scattering cross section over the mass
\begin{align}
	\sigma_{\rm DM-\nu} & \lesssim \frac{2.3\, m_{\rm DM}}{\Sigma_{\rm DM}}\;.
	\label{eq:criteria}
\end{align}
Since the flux attenuation at a fixed neutrino energy is exponentially suppressed with the optical depth of neutrinos in the DM medium, a more stringent criterion on the amount of attenuation would only relax our limits logarithmically as $\sim \mathrm{log({\Phi_{\rm em}}/{\Phi_{\rm obs}})}$.

\section{Bounds from the ultra-high energy event KM3-230213A}
\label{sec:resultsKM3Net}

The exact origin of event KM3-230213A is still not clear. However, a neutrino of galactic origin is disfavored because Galactic emission and known accelerators therein are insufficient to explain the event~\cite{KM3NeT:2025aps}. Then, a well-motivated explanation is that this neutrino belongs to the diffuse cosmogenic flux, produced by the same flux of cosmic rays with $E\gtrsim 100\, \mathrm{PeV}$ that is already measured \cite{PierreAuger:2023bfx}. While well-suited for explaining KM3-230213A, the absence of observations at IceCube and Auger --with a larger exposure-- places this scenario in a tension which ranges between 2.5$\sigma$ and 3.6$\sigma$~\cite{KM3NeT:2025ccp, KM3NeT:2025vut,Li:2025tqf}. Then, KM3-230213A would correspond to an upward fluctuation on the cosmogenic flux, presumably closely below the IceCube sensitivity limit. 

It has been argued that this tension can be slightly reduced to $2.0\sigma$ if the neutrino originated in a transient, localized, high-energy astrophysical source~\cite{Li:2025tqf, KM3NeT:2025ccp}. Up to 17 sources have been observed at different frequencies in the electromagnetic spectrum within the 99\% C.L. radius of the location of the event in the sky~\cite{KM3NeT:2025bxl}. One possible source of the KM3NeT event is the blazar PKS 0605-085~\cite{Dzhatdoev:2025sdi} as it is only $2.4^\circ$ from the central value of the position of the event, that is within the 2$\sigma$ region. Furthermore, it has been discussed that the event coincides with a $\gamma$-ray flare of the blazar, in analogy to the 2018 neutrino event observed from IceCube from TXS 0506+056, which was also associated with a gamma-ray flare \cite{IceCube:2018dnn}. 

Regardless of its origin, the flux of neutrinos will have gone through Milky Way's halo. 
The event's trajectory indicates that the neutrino flux does not go through or nearby the Galactic Center. While the column depth traversed, $\Sigma_{\mathrm{DM}}$, is not the largest possible in our galaxy, its value is robust against the choice of DM profile, as most models exhibit similar behavior at sufficiently large radii.
For the direction of KM3-230213A, we get $\Sigma_{\mathrm{DM}}^{\mathrm{(MW)}} \simeq 1.3\times 10^{22}$ GeV cm$^{-2}$. Then, Eq.~\eqref{eq:criteria} results in 
\begin{equation}
    \left.\dfrac{\sigma_{\rm DM-\nu}}{m_{\rm DM}}\right|_{\text {MW }}\lesssim 2.52\times10^{-22}\ \frac{{\rm cm}^2}{\rm GeV},
\end{equation}
at $E_\nu = 220^{+570}_{-110}\, \mathrm{PeV}$. 

More stringent bounds are expected if the neutrino was produced in the vicinity of a Supermassive Black Hole, e.g., in blazar PKS 0605-085. In that case, the propagation through the host galaxy is the dominant source of attenuation, see the Supplementary material for details.

We find that the column density in the spike of PKS 0605-058 is $\Sigma_{\text {DM }}^{\text {(spike)}} \simeq 4.4 \times 10^{28}$ GeV cm$^{-2}$, significantly larger than the intergalactic and MW contributions,
and results in the limit
\begin{equation}
    \left.\dfrac{\sigma_{\rm DM-\nu}}{m_{\rm DM}}\right|_{\text {spike}} \lesssim 5.2\times10^{-29}\ \frac{{\rm cm}^2}{\rm GeV},
\end{equation}
at $E_\nu = 220^{+570}_{-110}\, \mathrm{PeV}$. 

\begin{figure}[t]
\centering
\includegraphics[width=1\linewidth]{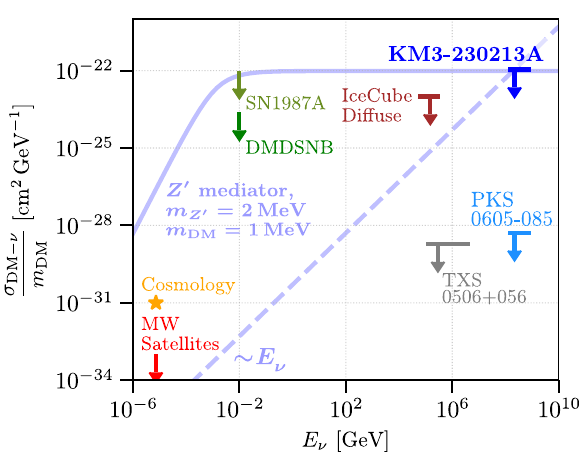}
    \caption{    Compilation of bounds on DM-neutrino scatterings across neutrino energies. A dark blue arrow indicates the exclusion limit from KM3-230213A assuming propagation only in the Milky Way, and lavender blue lines extrapolate it to lower energies, for $E_\nu^{-1}$ and a model with fermion DM and vector mediator, see Eq.~\eqref{eq:xs-fermionvector}. 
    A light blue arrow indicates the strengthened constraints if the emission arises from a blazar such as PKS 0605-085. Our constraints are compared to the typical cross section hinted by various cosmological observables like Ly-$\alpha$, $H_0$ and $S_8$ \cite{Hooper:2021rjc, DiValentino:2017oaw} (orange), constraints from Milky-Way Satellites \cite{Crumrine:2024sdn} (red), bounds from supernova SN1987A \cite{Mangano:2006mp, Chauhan:2025hoz} (light green), the diffused neutrino flux from DM-neutrino scatterings for galactic supernovae \cite{Chauhan:2025hoz} (dark green), bounds from TXS 0506+056 \cite{Ferrer:2022kei, Cline:2022qld, Zapata:2025huq} (grey), and from the diffuse flux of astrophysical neutrinos \cite{Arguelles:2017atb} (brown).}
    \label{fig:bounds_Edependent}
\end{figure}

Fig.~\ref{fig:bounds_Edependent} shows how KM3-230213A provides constraints on the DM-neutrino scattering cross section at an energy scale previously uncharted. In particular, in dark blue we show the constraints from attenuation only in the Milky Way, while in light blue we show the constraints from a DM spike at PKS 0605-085. 

While the bounds from TXS 0506+056 (grey) are apparently stronger than the ones from KM3-230213A and from PKS 0506+056, assuming this was the source of the neutrino, this need not be the case in models where the cross section rises with energy. In particular, the PKS 0506+056 bound becomes much stronger for $\sigma_{\mathrm{DM}-\nu}\sim E_\nu$. 

Bounds presented in this work are comparable to current MeV constraints \cite{Mangano:2006mp, Heston:2024ljf, Chauhan:2025hoz, Cappiello:2025tws, Manzari:2023gkt} for constant cross sections, but become stronger by orders of magnitude in models where the cross section rises with energy. 
Finally, we show how the Milky Way bound extrapolates to lower energies in a model of Dirac fermion DM with mass $m_{\mathrm{DM}}=1$ MeV and with a vector mediator of $m_{Z^{\prime}} = 2$ MeV.

\begin{figure*}[ht!]
\centering
\includegraphics[width=0.49\linewidth]{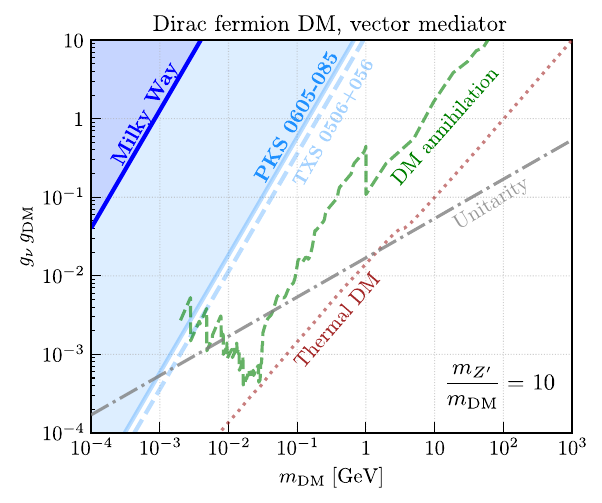}
\includegraphics[width=0.49\linewidth]{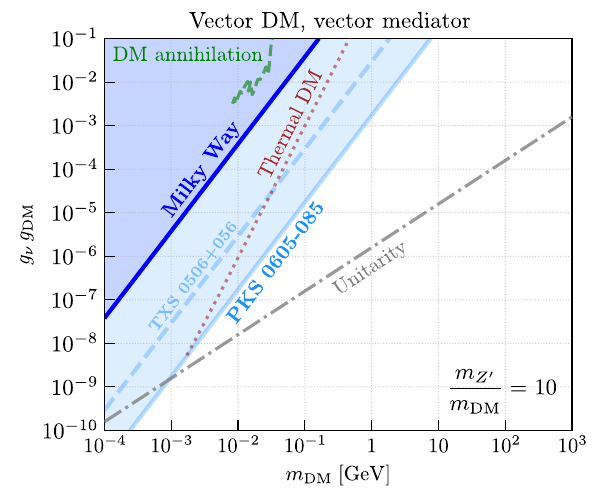}
    \caption{\textit{Left panel:} Upper limits on the product of gauge couplings of a new vector mediator to the DM Dirac fermion and neutrinos versus DM mass. We show the limits from KM3-230213A derived in this work as shaded contours. The dark blue contour consider attenuation only in the Milky Way, whereas light blue contour assumes the emission arises from PKS 0605-085. The light blue dashed lines indicates the limit from TXS~0506+056 \cite{Ferrer:2022kei, Cline:2022qld, Zapata:2025huq}. For comparison, we show the region of parameter space where the cross-section exceeds unitarity constraints (gray dot-dashed line), the observed relic abundance of DM can be achieved (red dotted line) and the limits on DM annihilation (green dashed line)~\cite{Arguelles:2019ouk}. \textit{Right panel:} Same as the left panel, but for the vector DM with a vector mediator case.}
    \label{fig:bounds_FermionVector}
\end{figure*}

\section{Bounds in simplified models of DM-neutrino interactions}
\label{sec:theo}
In this Section we reinterpret the previous constraints as bounds on the couplings of various simplified, but realistic, models of DM-neutrino interactions.

We consider left-handed neutrinos only and treat them as effectively massless. 
The coupling between DM and the mediator is denoted by $g_{\rm DM}$, and the coupling between neutrinos and the mediator by $g_\nu$.
In this Section, cross sections are written in the high-energy limit, $E_\nu\gg m_{\rm DM}, M_{Z^{\prime}}$. The Supplementary Material presents all intermediate calculations and full expressions for the $\nu\, \rm DM\to\nu\,\rm DM$ cross section, for the models studied in this section together with calculations and results for additional simplified models.

In these simplified models, demanding that the observed relic abundance of DM in the Universe is fulfilled relates their couplings and masses.
For the standard freeze-out mechanism, this can be done by means of the thermally-averaged annihilation cross section. In order to reproduce the DM of the Universe, the thermally averaged cross section should approximately be  $\langle \sigma v \rangle_{\textrm{freeze-out}} \simeq (2-5)\times 10^{-26} \mathrm{~cm}^3 / \mathrm{s}$, and we pick as a benchmark \cite{Steigman:2012nb}
\begin{equation}
\langle \sigma v \rangle_{\textrm{freeze-out}} \simeq 4.4 \times 10^{-26} \mathrm{~cm}^3 / \mathrm{s}\, .
\end{equation}
In addition to the limits presented, some of our simplified models will feature new neutrino self-interactions (SI). These SI have been extensively searched for and several limits might apply, see~\cite{Berryman:2022hds}.

\subsection{Fermion DM with vector mediator}

We first consider a fermionic DM candidate, $\chi$, that interacts with neutrinos through a vector mediator, $Z^{\prime}$, as it arises naturally in $U(1)$ extensions of the Standard Model~\cite{Holdom:1985ag,Fox:2008kb,Bell:2014tta}.  
On the one hand, if $\chi$ is a Dirac fermion, 
the Lagrangian of the interaction will be 
\begin{equation}\label{eq:lag_fermion_vector}
\mathcal{L} =\ -g_{\rm DM} \bar{\chi} \gamma^\mu Z^\prime_{\mu}\chi - g_\nu \overline{\nu_L} \gamma^\mu Z^\prime_\mu \nu_L  + \text{h.c.},
\end{equation}
where we assume that $g_{\rm L, DM}=g_{\rm R, DM}=g_{\rm  DM}$. 
On the other hand, if $\chi$ is a Majorana fermion we got
\begin{equation}
\mathcal{L} =\ -\frac{g_{\rm DM}}{2} \bar{\chi} \gamma^\mu \gamma_5\,Z^\prime_{\mu}\chi - g_\nu \overline{\nu_L} \gamma^\mu Z^\prime_\mu \nu_L  + \text{h.c.}\, .
\end{equation}
In the high-$E_\nu$ limit, the Dirac cross section is
%
\begin{align}\label{eq:xs-fermionvector}
  \sigma_{\rm DM-\nu} \approx \,  &\frac{g_{\text{DM}}^2\, g_\nu^2}{8\, \pi} \Bigg[\frac{1}{ M_{Z'}^2}\\ 
&+ \frac{1}{2\, m_{\text{DM}}\, E_\nu}
\left( \log\left( \frac{M_{Z'}^2}{2\, m_{\text{DM}}\, E_\nu} \right) + \frac{1}{2} \right)\Bigg]\, ,\nonumber
\end{align}
%
while the Majorana cross section carries an extra $\frac{1}{4}$ factor. 
As shown in the Supplementary Material, at low energies the cross sections scale with $E_\nu$, while at high energies they approach a constant value. 
If we have a DM candidate in the TeV range and a mediator a few orders of magnitude higher we can escape the high-energy limit as $M_{Z^\prime}\sim E_\nu$.
Even in this case, $\sigma_{\mathrm{DM}-\nu}$ is then 
suppressed by a $M_{Z^\prime}^{-4}$ factor.

In the Dirac scenario, the thermally-averaged annihilation cross section can be approximated as \cite{Olivares-DelCampo:2017feq}
\begin{equation}\label{eq:relic_fermionvector}
\langle \sigma v \rangle \simeq \frac{g_{\rm DM}^2 g_\nu^2}{2 \pi} \frac{m_{\mathrm{DM}}^2}{\left(4 m_{\mathrm{DM}}^2-M_{Z^{\prime}}^2\right)^2} \, .
\end{equation}
Contrarily, Majorana DM annihilates as a \textit{p}-wave process, and therefore $\langle \sigma v \rangle$ is proportional to $v_{\mathrm{CM}}^2$, the squared center-of-mass velocity. Fig.~\ref{fig:bounds_FermionVector} shows the KM3-230213A constraints on $g_\nu g_{\mathrm{DM}}$ 
versus $m_{\rm DM}$, for Dirac DM with fixed 
$m_{Z^{\prime}}/m_{\rm DM}=10$. Bounds for Majorana DM are weakened by a factor of 2, see the Supplementary Material. For $M_{\mathrm{Z}'}$ fixed, bounds from high-energy neutrino attenuation scale roughly as $m_{\mathrm{DM}}^{-1}$. Bounds also get stronger for lighter mediators. In this model in particular, $m_{Z^{\prime}} < m_{\rm DM}$ is allowed since it does not affect DM stability through decay.

Most importantly, Fig.~\ref{fig:bounds_FermionVector} shows that the couplings probed in this simplified model correspond for $m_{\rm DM} \gtrsim 1\, {\rm MeV}$ to cross sections which exceed partial wave unitarity i.e., $\sigma_{\rm DM-\nu} > 4\pi/k^2$, with $k^2$ the center-of-mass momentum squared at $E_{\nu}=220\, {\rm PeV}$. The constraint on $g_\nu g_{\rm DM}$ must then be corrected by a concrete UV completion which regularizes this simplified model. A model-independent bound on the high-energy behaviour of the cross-sections calculated is the so called Froissart-Martin bound~\cite{Froissart:1961ux, Martin:2013xcr}, which is less stringent than the partial wave unitarity bound shown in the Figure. 
Altogether, this makes the important conclusion that, in order for the constraints on $\sigma_{\rm DM-\nu}$ to be meaningful, models must be more complex than these presented here.

Finally, we further confront these bounds with complementary bounds from DM-annihilation in the galactic halo \cite{Arguelles:2019ouk}, which will be weaker for the Majorana case due to its \textit{p}-wave nature. These bounds 
only apply if the DM is a Majorana fermion, or if the DM of the Universe remains symmetric up until today. For completeness, we show how the bounds for this model do not yet approach the parameter space where the relic abundance of the Universe can be achieved via freeze-out, in a red dotted line. 

\subsection{Vector DM with vector mediator}

A more exotic case is the one including a vector DM candidate with another vector mediating the interaction with neutrinos. This type of interactions could arise with Abelian symmetries with Chern-Simons like interactions; or with non-Abelian symmetries such as new $SU(2)$~\cite{Belyaev:2022shr}. We consider the Lagrangian
\begin{equation}
\mathcal{L} =\ -\frac{1}{2}g_{\rm DM} V^\nu \partial_\nu V^\mu Z^\prime_{\mu} - g_\nu \overline{\nu_L} \gamma^\mu Z^\prime_\mu \nu_L  + \text{h.c.},
\end{equation}
where $V$ is our DM candidate. The high-energy cross section grows with $E_\nu$,
%
\small
\begin{align}
    \sigma_{\rm DM-\nu} &\approx \frac{g_{\text{DM}}^2\, g_\nu^2}{3072\, \pi\, m_{\text{DM}}^2}
\Bigg[\dfrac{2\,E_\nu}{m_{\text{DM}}} \\ & -
\bigg(8  - 4 \dfrac{M_{Z'}^2}{m_{\text{DM}}^2}\bigg) \log\Bigg( \frac{M_{Z'}^2}{m_{\text{DM}}^2 y} \Bigg)
+ 6 \dfrac{M_{Z'}^2}{m_{\text{DM}}^2} - 15
\Bigg]\, . \nonumber
\end{align}
\normalsize
Here the DM opacity is significant for neutrinos with energies as high as the KM3-230213A event. As we can see in Fig. \ref{fig:bounds_FermionVector}, bounds are competitive for masses below the $\sim$ GeV, as the cross section scales with $m_{\rm DM}^{-2}$. 
In this model, the annihilating cross section is
\begin{equation}\label{eq:relic_vectorvector}
\langle \sigma v \rangle \simeq \frac{g_{\rm DM}^2 g_\nu^2}{\pi} \frac{m_{\mathrm{DM}}^2}{\left(4 m_{\mathrm{DM}}^2-M_{Z^{\prime}}^2\right)^2}v_{\rm CM}^2\, ,
\end{equation}
down to masses around the MeV. In this mass range, the Milky Way limit from KM3-230213A is close to probing the parameter space that explains the DM relic abundance, but vector DM can be lighter if we go beyond the thermal freeze-out mechanism. It can be produced via freeze-in down to keV masses~\cite{Bernal:2015ova}, with the misalignment mechanism for eV masses and below~\cite{Nelson:2011sf}, or even from gravitational production in a wide range of masses~\cite{Ema:2019yrd}. As in the previous scenario, this bound exceeds unitarity.

\section{Discussion}
\label{sec:discussion}

In this work, we have leveraged the detection of a high-energy event by the KM3NeT collaboration to constrain DM-neutrino interactions, under the assumption that the event corresponds to a high-energy neutrino of extragalactic origin.

For non-zero DM-neutrino interactions, neutrinos would be deflected and down scattered when traversing DM halos, effectively attenuating the flux reaching the Earth. Assuming that the flux responsible for the high-energy event has been attenuated less than an one order of magnitude when traversing the Milky Way, we constrain the ratio of the cross section of these interactions and the mass of the DM to be $\sigma_{\rm DM-\nu}/m_{\rm DM} \lesssim 10^{-22}$~cm$^{2}$~GeV$^{-1}$. If the high-energy neutrino had been produced in a host galaxy in a region of high DM density, e.g., the blazar PKS 0605-085; the constraint becomes $\sim$~6 orders of magnitude stronger.

The limits here derived belong to an unexplored energy range and therefore, it is relevant to translate them into constraints on specific--yet simplified-- models featuring DM-neutrino interactions. Models with a Dirac fermion or scalar DM and a vector mediator predict a constant cross section at high energies, and thus limits from KM3-230213A are only stringent as others coming from cosmological and laboratory probes if the high-energy neutrino is produced in the vicinity of a SMBH. However, for Majorana DM and a vector mediator the limits from attenuation in the Milky Way DM halo are stronger the bounds from DM annihilation in some regions of parameter space, since the latter are p-wave suppressed.
Furthermore, models with vector DM and mediator predict a cross section which increases with the neutrino energy. These models yield limits from solely attenuation in the Milky Way halo which improve DM annihilation bounds by orders of magnitude. Importantly, limits from attenuation allow us to probe the region of parameter space of this model where Dark Matter is thermally produced in the early Universe. Still, if $m_{\rm DM}\gtrsim 1\, {\rm MeV}$ and $m_{Z^{\prime}}/m_{\rm DM}=10$, our bounds exceed unitarity, and therefore these simplified models must be corrected by an UV completion which regularizes them. This motivates the study of more complex dark sectors which could explain neutrino absorption by DM in the ultra-high-energy range. 

In the future, high-energy neutrinos in association with a source or arriving in a direction closer to the galactic center would further improve these results since the associated DM column density could be orders of magnitude larger. In the advent of multimessenger astronomy and the increasing sensitivity to very high-energy neutrinos, this work highlights the relevance of astrophysical neutrinos to probe DM-neutrino interactions.

\begin{acknowledgments}
We are grateful to Mar Císcar-Monsalvatje, Salvador Centelles Chuliá, Jorge Martínez Vera and Neus Penalva for useful discussions. We have used FeynCalc \cite{Mertig:1990an,Shtabovenko:2016sxi,Shtabovenko:2020gxv, Shtabovenko:2023idz} to perform the calculations presented in the Supplementary Material.  TBM acknowledges support from the Spanish MCIN/AEI/10.13039/501100011033 grant PID2022-126224NB-C21 and the grant PRE2020-091896, the European Union’s Horizon 2020 research and innovation program under the Marie Skłodowska-Curie grants HORIZON-MSCA-2021-SE-01/101086085-ASYMMETRY and H2020-MSCA-ITN-2019/860881-HIDDeN, and support from the ``Unit of Excellence Maria de Maeztu 2020-2023'' award to the ICC-UB CEX2019-000918-M. The work of GH is supported by the U.S. Department
of Energy under award number DE-SC0020262. PMM has received support from the Villum
Foundation (Project No. 13164, PI: I. Tamborra). JTC acknowledges support by the research grant TAsP (Theoretical Astroparticle Physics) funded by Istituto Nazionale di Fisica Nucleare (INFN) and the research grant ``Addressing systematic uncertainties in searches for DM No.
2022F2843'' funded by MIUR.
\end{acknowledgments}

\appendix
\section{Column density of dark matter}
Neutrinos are attenuated at the host galaxy where they are produced (if any), in the intergalactic medium, and in the Milky Way. Here we compute the column density of dark matter traversed by neutrinos in these three environments.

\begin{figure*}[ht!]
\centering
\includegraphics[width=0.65\linewidth]{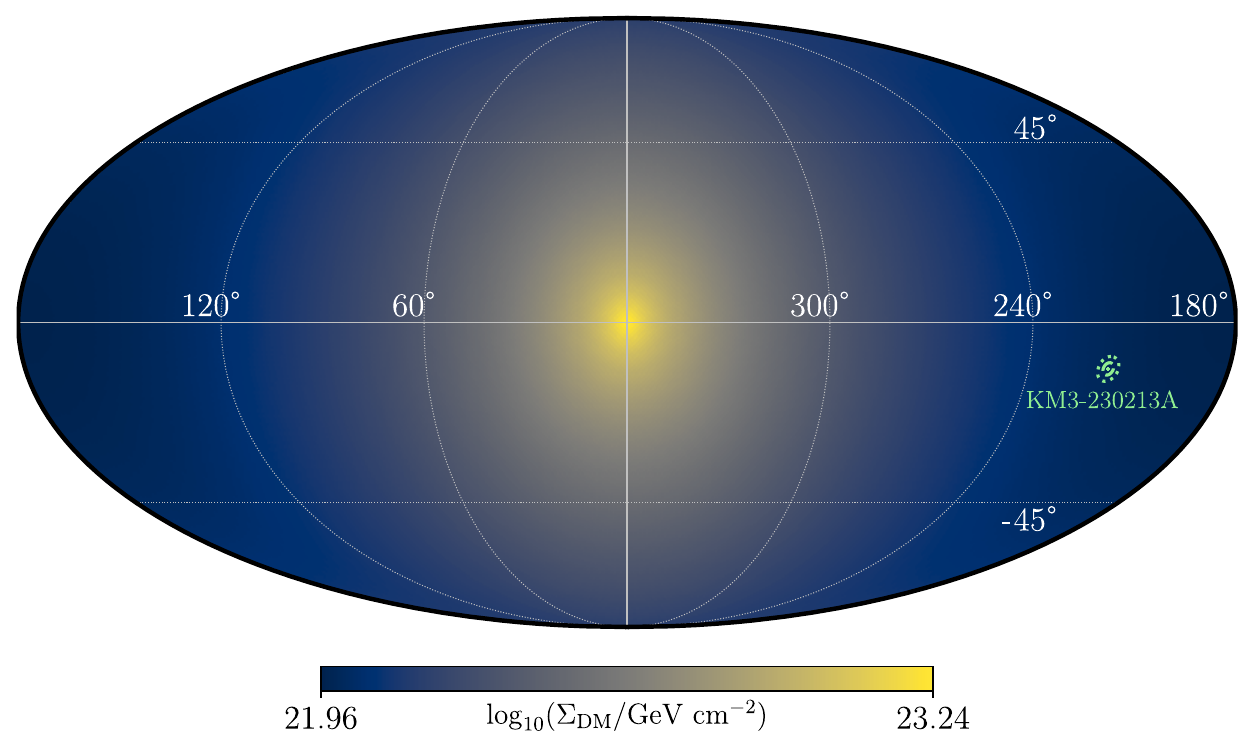}
     \caption{Sky map of the DM column density of the Milky Way's halo, $\Sigma_{\rm DM}$, for the DM profile used in this work, together with the location of KM3-230213A event and the 99\% (dotted) and 68\% (dashed) uncertainty regions.}
     \label{fig:Map}
\end{figure*}

We first start describing the dark matter column density from the Milky Way. This contribution is guaranteed, provided the neutrino is of extragalactic origin. It can be computed as

\begin{equation}
    \Sigma_{\rm DM}(b,l) = \int_{0}^{r(b,l)}\rho_{\mathrm{DM}}\left(r^\prime(b,l)\right)d{ r^\prime} = \Sigma_{\mathrm{DM}}(b,l)\, .\label{eq:ColumnDensity}
\end{equation}
Here, $b$ and $l$ are the galactic latitude and longitude respectively, $r$ is the distance along the line of sight and $m_{\rm DM}$ is the DM mass. In the main text we make use of the DM column density,  $\Sigma_{\mathrm{DM}}$. In this work, we use a Navarro-Frenk-White (NFW) profile~\cite{Navarro:1995iw,Navarro:1996gj} with parameters $r_s = 14.46$ kpc and $\rho_s = 0.566$ GeV/cm$^3$~\cite{Cirelli:2024ssz}.
For the direction of KM3-230213A, we then get $\Sigma_{\mathrm{DM}}^{\mathrm{(MW)}} \simeq 1.3\times 10^{22}$ GeV/cm$^2$. We further show the column density obtained for different regions in the Sky in Figure \ref{fig:Map}, highlighting the direction of KM3-230213A.

We discuss now the column density contribution arising from the intergalactic medium. The column density of intergalactic dark matter encountered by a particle emitted at a source with redshift $z$, along the line of sight between the source and the Earth (LOS) is

\begin{equation}
\int_{\text {\rm LOS}} \rho_{\rm DM}(z) d l=\int \rho_{\rm DM}(z) \frac{c d t}{d z} d z
\end{equation}

where $\rho_{\rm DM}(z)=\rho_{\rm DM}(z=0)(1+z)^3, $ $d t / d z=-((1+z)\, H(z))^{-1}$ and $H(z)=$ $H_0 \sqrt{\Omega_{\Lambda}+\Omega_m(1+z)^3}$. We take $H_0=67.4 \mathrm{~km} / \mathrm{sec} / \mathrm{Mpc}, \Omega_{\Lambda}=0.685$, and $\Omega_m=0.315$ \cite{Planck:2018vyg}. Furthermore,
\begin{equation}
\rho_{\rm DM}(z=0)=\Omega_{\rm DM}\, \rho_{\rm crit}\simeq 1.24 \times 10^{-6}\, \mathrm{GeV/cm}^3
\end{equation}
Under this prescription, if neutrinos arised from a blazar such as PKS 0605-085 ($z=0.87$), the corresponding cosmological column density of dark matter would be $\Sigma_{\rm DM}^{\rm PKS} \simeq 3.3 \times 10^{22}$GeV/cm$^2$. This value is comparable to the column density expected in the Milky Way halo, thus relevant.

Finally, we are left to consider the attenuation of neutrinos in the halo of the host galaxy where they may have been produced. As previously mentioned, the redshift of PKS 0605-085 has been estimated as $z=0.87$ \cite{Shaw_2012}, allowing us to infer a most likely black hole mass of $M_{\rm BH} \sim 10^{9} M_{\odot}$~\cite{Vestergaard:2009ii,Caramete:2009nh}. The emission of high-energy neutrinos from the blazar TXS 0506+056 likely occurs at the Broad Line Region \cite{Padovani_2019}, and for PKS 0605-085 we will work under this assumption. That is, $R_{\rm em} \sim 10^{4} R_S$, with $R_S$ the Schwarzschild radius of the central black hole. The column density of DM particles in the surrounding DM spike around the black hole is \cite{Ferrer:2022kei}
\begin{equation}
\Sigma_{\text {DM }}^{\text {(spike)}} \simeq \frac{\rho_{\mathrm{sp}}\left(R_{\mathrm{em}}\right) R_{\mathrm{em}}}{\left(\gamma_{\mathrm{sp}}-1\right)}\left[1-\left(\frac{R_{\mathrm{sp}}}{R_{\mathrm{em}}}\right)^{1-\gamma_{\mathrm{sp}}}\right]\, ,
\end{equation}
with the spike density profile $\rho_{\rm sp}$ given by \cite{Gondolo_1999}
\begin{equation}
\rho_{\mathrm{sp}}(r)=\rho_R\,  g_\gamma(r)\left(\frac{R_{\rm sp}}{r}\right)^{\gamma_{\mathrm{sp}}}
\end{equation}
with the size of the spike $R_{\mathrm{sp}}=\alpha_\gamma r_0\left(M_{\mathrm{BH}} /\left(\rho_0 r_0^3\right)^{\frac{1}{3-\gamma}}\right)$. We consider an initial NFW profile with $\gamma=1$, such that $\alpha_\gamma \approx 0.1$, and $\gamma_{\mathrm{sp}}=\frac{9-2 \gamma}{4-\gamma}=7/3$.  $g_\gamma(r) \simeq\left(1-\frac{4 R_S}{r}\right)$, with $R_S$ the Schwarzschild radius, and $\rho_{\mathrm{R}}$ is a normalization factor, chosen to match the density profile outside of the spike, $\rho_R=\rho_0\left(R_{s p} / r_0\right)^{-\gamma}$. We consider a scale radius $r_0=10$ kpc. Finally, the normalization $\rho_0$ is numerically fixed such that the integrated DM profile in the Galaxy yields a halo mass of $M_{\rm DM} \simeq 10^{13} M_{\odot}$. Under this prescription, the column density in the spike of PKS 0605-058 is $\Sigma_{\text {DM }}^{\text {(spike)}} \simeq 4.4 \times 10^{28}$ GeV/cm$^2$.

\section{Detailed DM - neutrino cross section calculations}
In this section of the Supplementary Material, we detail the calculations of the cross section for several simplified models including the ones discussed in section IV and expand on other simplified models, showing the results for those which do not have an energy suppression on the cross-section. We consider scalar, fermionic (both Dirac and Majorana), and vector DM candidates, as well as scalar, fermionic, and vector mediators. 
We denote spin-0 particles by $\phi$, spin-$\frac{1}{2}$ particles by $\chi$, and spin-1 particles by $Z^\prime$ (or $V$ if two vectors are involved). The corresponding mediator masses are denoted by $M_\phi$, $M_\chi$, and $M_{Z'}$, while the DM candidate mass is represented by $m_{\rm DM}$, regardless of the particle type. We are in agreement with the cross sections calculated in \cite{Olivares-DelCampo:2017feq} up to factors of $2$ which origin we cannot identify. Here we detail the amplitudes considered, their spin averaged square and the final result for each cross section, as well as the high-energy limit, which is the relevant one for our work.

\label{sec:appendix_xsect}
\subsection{Kinematic Relations}

For clarity, we consider particle 1 the incoming neutrino with momentum $p_1$, particle 2 the incoming DM particle with momentum $p_2$, particle 3 the outgoing neutrino with momentum $k_1$ and particle 4 the outgoing DM particle with momentum $k_4$. We start by considering the kinematic conditions and relations for scattering processes~\cite{ParticleDataGroup:2024cfk}:

\begin{align}
&m_{\nu}  = m_1 = m_3 = 0, \qquad m_{\rm DM}  = m_2 = m_4, \\
&p_1^2 = 0, \quad p_2^2 = m_{\rm DM}^2, \qquad 
k_1^2 = 0, \quad k_2^2 = m_{\rm DM}^2.
\end{align}

Using the mandelstam variables
\begin{align}   
    &s + t + u  = 2m_{\rm DM}^2,
\end{align}

and the dot products
\begin{align}
p_1 \cdot p_2 &= k_1 \cdot k_2 = \dfrac{1}{2} \left(s - m_{\rm DM}^2\right), \\
p_1 \cdot k_1 &= p_2 \cdot k_2 = -\dfrac{1}{2} t, \\
p_1 \cdot k_2 &= p_2 \cdot k_1 = \dfrac{1}{2} \left(m_{\rm DM}^2 - u\right).
\end{align}

\subsubsection{Energy Expressions}

\begin{align}
E^{\rm CM}_{p_1} = E^{\rm CM}_{k_1} =\dfrac{s - m_{\rm DM}^2}{2\, \sqrt{s}} \qquad
E^{\rm CM}_{p_2} = E^{\rm CM}_{k_2} =\dfrac{s + m_{\rm DM}^2}{2\, \sqrt{s}}
\end{align}

\subsubsection{Cross section}
Integration limits in the case of a massless incoming particle:
\[
\dfrac{d\sigma}{dt} = \dfrac{1}{16\pi\left(s - m_{\rm DM}^2\right)^2}  \abs{\mathcal{M}(t,s)}^2,
\]

with integration limits

\[
t_0 = 0 , \qquad t_1 = -\dfrac{\left(s - m_{\rm DM}^2\right)^2}{s}.
\]

One can also work with the mandelstam variable $u$ for simplicity in the processes with \textit{u-channel} diagrams using 
\[
\dfrac{d\sigma}{du} = -\dfrac{1}{16\pi\left(s - m_{\rm DM}^2\right)^2}  \abs{\mathcal{M}(u,s)}^2,
\]
with integration limits
\[
u_0 = 2\, m_{\rm DM}^2 - s , \qquad u_1 = \dfrac{m_{\rm DM}^4}{s}.
\]

Since we will be dealing with a high-energy incoming neutrino,  we can work with the variable
\[
y =  \dfrac{s}{m_{\rm DM}^2} - 1,
\]
so we can expand the cross sections in powers of $1/y$ to get the high-energy limit and in powers of $y$ to get the low-energy behavior. The variable $y$ is proportional to the incoming neutrino energy, $E_\nu$, in the LAB frame where the DM particle is at rest.

\subsection{Case 1.1: Dirac Fermion DM with Vector Mediator}

The interaction Lagrangian is:\footnote{Here we assume that $g_{\rm L, DM}=g_{\rm R, DM}=g_{\rm  DM}$}
\begin{equation}
\mathcal{L} =\ -g_{\rm DM} \bar{\chi} \gamma^\mu Z^\prime_{\mu}\chi - g_\nu \overline{\nu_L} \gamma^\mu Z^\prime_\mu \nu_L  + \text{h.c.}
\end{equation}

The amplitude of neutrino scattering on DM is
\begin{widetext}
\begin{align}
\mathcal{M}_{\text{t}(t,s)} =\, &g_{\rm DM}\, g_\nu\, \bar{u}(k_1) \gamma^\mu P_L\, u(p_1) \frac{g_{\mu\nu}-\frac{(p_1 - k_1)_\mu(p_1 - k_1)_\nu}{M_{Z^\prime}^2}}{t - M_{Z^\prime}^2} \bar{u}(k_2) \gamma^\nu P_L\, u(p_2)\nonumber \\
&+ g_{\rm DM}\, g_\nu\, \bar{u}(k_1) \gamma^\mu P_L\, u(p_1) \frac{g_{\mu\nu}-\frac{(p_1 - k_1)_\mu(p_1 - k_1)_\nu}{M_{Z^\prime}^2}}{t - M_{Z^\prime}^2} \bar{u}(k_2) \gamma^\nu P_R\, u(p_2) \,.
\end{align}
\end{widetext}
The spin-averaged amplitude squared reads
\begin{align}
\overline{\abs{\mathcal{M}(t,s)}^2} =\dfrac{g_{\rm DM}^2\, g_\nu^2 \Big(2\left(m_{\rm DM}^2 - s \right)^2+ 2\,s\,t + t^2\Big)}{\left(t - M_{Z^\prime}^2\right)^2} \, ,
\end{align}
and therefore, the total cross section is 
\begin{align}\label{eq:xsectDiracDMVector}
    \sigma_{\rm DM-\nu} = & \frac{g_{\text{DM}}^2\, g_\nu^2}{16\, \pi\, m_{\text{DM}}^2} \nonumber \\
        &\Bigg[ \frac{M_{Z'}^2}{m_{\text{DM}}^2\, y^2 + M_{Z'}^2 (y + 1)}   +\frac{2 m_{\text{DM}}^2}{M_{Z'}^2}        \nonumber \\
        & + \frac{1}{y + 1} -2 \dfrac{\left( m_{\text{DM}}^2 (y + 1) + M_{Z'}^2 \right)}{m_{\text{DM}}^2 y^2}
         \nonumber \\
        & \log\left(  \dfrac{m_{\text{DM}}^2 y^2}{ M_{Z'}^2(y + 1)} + 1  \right)
         \Bigg]\,.
\end{align}

In the high-energy limit, it reads
\begin{align}
   \sigma_{\rm DM-\nu} \approx \, &\frac{g_{\text{DM}}^2\, g_\nu^2}{8\, \pi} \Bigg[\frac{1}{ M_{Z'}^2} \nonumber \\ &
+ \frac{1}{ m_{\text{DM}}^2\, y}
\left( \log\left( \frac{M_{Z'}^2}{m_{\text{DM}}^2\, y} \right) + \frac{1}{2} \right)\Bigg]
+ \mathcal{O}\left( \frac{1}{y^2} \right)\,,
\end{align}
and is constant with energy for $y\gg (m_{Z^{\prime}}/m_{\rm DM})^{2}$, i.e., for high energies. If we go to the low-energy limit, $y\ll (m_{Z^{\prime}}/m_{\rm DM})^{2}$, we get
\begin{align}
   \sigma_{\rm DM-\nu} \approx \, &\frac{g_{\text{DM}}^2\, g_\nu^2}{16\, \pi} \dfrac{y^2\, m_{\text{DM}}^2}{M_{Z'}^4}
+ \mathcal{O}\left( y^3 \right)\,,
\end{align}
which grows as $y^2\propto E_\nu^2$.

\subsection{Case 1.2: Majorana Fermion DM with Vector Mediator}

The interaction Lagrangian is
\begin{equation}
\mathcal{L} =\ -\frac{g_{\rm DM}}{2} \bar{\chi} \gamma^\mu \gamma_5\,Z^\prime_{\mu}\chi - g_\nu \overline{\nu_L} \gamma^\mu Z^\prime_\mu \nu_L  + \text{h.c.}
\end{equation}
and the scattering for elastic neutrino scattering on DM is
\begin{widetext}
\begin{align}
\mathcal{M}_{\text{t}}(t,s) =\, &\frac{g_{\rm DM}}{2} \, g_\nu\, \bar{u}(k_1) \gamma^\mu P_L\, u(p_1) \frac{g_{\mu\nu}-\frac{(p_1 - k_1)_\mu(p_1 - k_1)_\nu}{M_{Z^\prime}^2}}{t - M_{Z^\prime}^2} \bar{u}(k_2) \gamma^\nu \gamma_5\, u(p_2) \,.
\end{align}
\end{widetext}

Then, the spin-averaged amplitude squared reads
\begin{align}
\overline{\abs{\mathcal{M}(t,s)}^2} =&\dfrac{g_{\rm DM}^2 g_\nu^2 }{4\left(t - M_{Z^\prime}^2\right)^2}\nonumber \\ &\Big(2\,m_{\rm DM}^4 - 4\, m_{\rm DM}^2\, \left(s+t \right)+ 2\, s^2 + 2\,s\,t + t^2\Big)
\end{align}
and the total cross section is given by
\begin{align}\label{eq:xsectMajoranaDMVector}
    \sigma_{\rm DM-\nu} = & \frac{g_{\text{DM}}^2\, g_\nu^2}{64\, \pi\, m_{\text{DM}}^2} \nonumber \\
        &\Bigg[ \frac{M_{Z'}^2 - 4 m_{\text{DM}}^2} {m_{\text{DM}}^2\, y^2 + M_{Z'}^2 (y + 1)} + \frac{2 m_{\text{DM}}^2}{M_{Z'}^2}  \nonumber \\
        &  + \frac{1}{y + 1}   -2 \dfrac{\left( m_{\text{DM}}^2 (y - 1) + M_{Z'}^2 \right)}{m_{\text{DM}}^2 y^2}
       \nonumber \\
        & \log\left(  \dfrac{m_{\text{DM}}^2 y^2}{ M_{Z'}^2(y + 1)} + 1  \right)
         \Bigg]\,.
\end{align}

\begin{figure*}[ht!]
\centering
\includegraphics[width=0.49\linewidth]{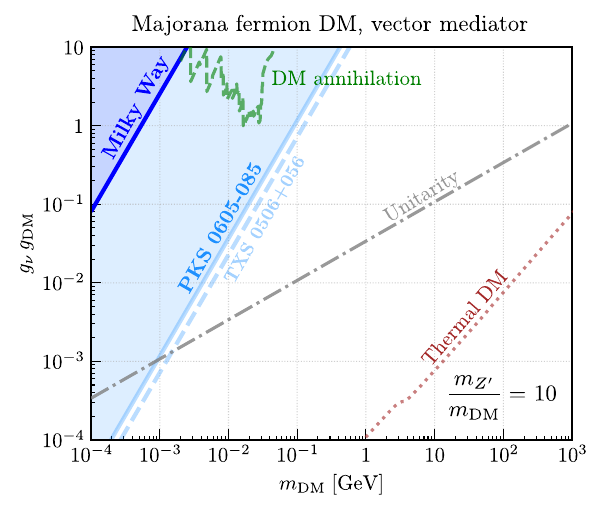}
\includegraphics[width=0.49\linewidth]{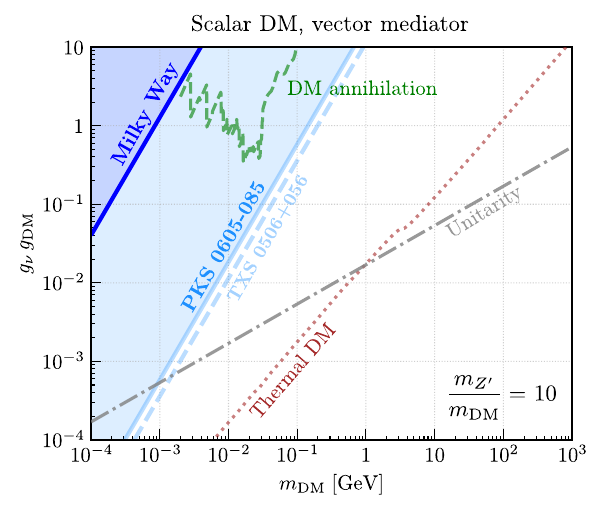}
    \caption{Upper limits on the product of gauge couplings of a new vector mediator to the DM Majorana fermion and neutrinos (left) and on the product of gauge couplings of a new vector mediator to the DM scalar and neutrinos (right) versus DM mass. We show the limits from KM3-230213A derived in this work as shaded contours. The dark blue contour consider attenuation only in the Milky Way, whereas light blue contour assumes the emission arises from PKS 0605-085. The light blue dashed lines indicates the limit from TXS~0506+056 \cite{Ferrer:2022kei, Cline:2022qld, Zapata:2025huq}.For comparison, we show the region of parameter space where the cross-section exceeds unitarity constraints (gray dot-dashed line), the observed relic abundance of DM can be achieved (red dotted line) and the limits on DM annihilation (green dashed line)~\cite{Arguelles:2019ouk}.}
    \label{fig:bounds_Majoranavector}
\end{figure*}

In the high-energy limit, the cross section reads
\begin{align}
   \sigma_{\rm DM-\nu} \approx \, &\frac{g_{\text{DM}}^2\, g_\nu^2}{32\, \pi} \Bigg[\frac{1}{ M_{Z'}^2} \nonumber \\
        & 
+ \frac{1}{ m_{\text{DM}}^2\, y}
\left( \log\left( \frac{M_{Z'}^2}{m_{\text{DM}}^2\, y} \right) + \frac{1}{2} \right)\Bigg]
+ \mathcal{O}\left( \frac{1}{y^2} \right)\,,
\end{align}
and is constant with energy for $y\gg (m_{Z^{\prime}}/m_{\rm DM})^{2}$, i.e., for high energies.
If we go to the low-energy limit, $y\ll (m_{Z^{\prime}}/m_{\rm DM})^{2}$, we get
\begin{align}
   \sigma_{\rm DM-\nu} \approx \, &\frac{3\, g_{\text{DM}}^2\, g_\nu^2}{64\, \pi} \dfrac{y^2\, m_{\text{DM}}^2}{M_{Z'}^4}
+ \mathcal{O}\left( y^3 \right)\,,
\end{align}
which grows as $y^2\propto E_\nu^2$. The bounds for this model are presented in the left panel of Fig.~\ref{fig:bounds_Majoranavector}. We see that attenuation bounds are very similar to Dirac DM, see Fig. 3 of the main text, but DM annihilation limits get much weaker due to the \textit{p}-wave nature of the process.

\subsection{Case 2.1: Dirac Fermion DM with Scalar Mediator}

The interaction Lagrangian is
\begin{equation}
\mathcal{L} = -g \overline{\chi_R}\, \phi\, \nu_L + \text{h.c.}
\end{equation}
and the scattering amplitude of interest is
\begin{equation}
\mathcal{M}_u(u,s) = g^2 \bar{u}(k_1) P_R u(p_2)\dfrac{1}{u - M_\phi^2} \bar{u}(k_2) P_L u(p_1)\,.
\end{equation}

The spin-averaged amplitude squared reads
\begin{align}
    \overline{\abs{\mathcal{M}(u,s)}^2} = &\dfrac{g^4 \left(m_{\rm DM}^2 - u\right)^2}{4 \left(u - M_\phi^2\right)^2} 
\end{align}
and hence, the total cross section is
\begin{widetext}
\begin{align}
    \sigma_{\rm DM-\nu} =   &\frac{g^4}{64\, \pi\, m_{\text{DM}}^4\, y^2}  \Bigg[2 (m_{\text{DM}}^2 - M_\Phi^2)\log\left(\dfrac{m_{\text{DM}}^2 (y - 1) + M_\Phi^2}{M_\Phi^2 - \frac{m_{\text{DM}}^2}{y + 1}} \right) \nonumber \\
    &+ \frac{m_{\text{DM}}^2\, y^2\Big(  2 m_{\text{DM}}^4
    + m_{\text{DM}}^2\, M_\Phi^2 \left((y - 2)\,y - 4\right) + 2 M_\Phi^4 (y + 1) \Big)}{(y + 1) \left(m_{\text{DM}}^2 (y - 1) + M_\Phi^2\right)
    \left(M_\Phi^2 (y + 1) - m_{\text{DM}}^2\right)}   \Bigg]\,.
\end{align}
\end{widetext}

In the high-energy limit, the cross section is then
\begin{align}
    \sigma_{\rm DM-\nu} &\approx \frac{g^4}{64\, \pi\, m_{\text{DM}}^2\, y} 
    + \mathcal{O}\left(\frac{1}{y^2}\right) \,,
\end{align}
supressed by $y^{-1}\propto E_\nu^{-1}$ for $y\gg (m_{Z^{\prime}}/m_{\rm DM})^{2}$, i.e., for high energies.
If we go to the low-energy limit, $y\ll (m_{Z^{\prime}}/m_{\rm DM})^{2}$, we get
\begin{align}
   \sigma_{\rm DM-\nu} \approx \, &\frac{g^4}{64\, \pi\,}\dfrac{y^2\,m_{\text{DM}}^2}{\left(m_{\text{DM}}^2-m_{\Phi}^2\right)^2} 
+ \mathcal{O}\left( y^3 \right)\,,
\end{align}
which grows as $y^2\propto E_\nu^2$.

\subsection{Case 2.2: Majorana Fermion DM with Scalar Mediator}

The interaction Lagrangian is
\begin{equation}
\mathcal{L} = -g \overline{\chi_R}\, \phi\, \nu_L + \text{h.c.}
\end{equation}
and the amplitude for neutrino elastic scattering on DM is given by
\begin{align}
\mathcal{M}_u(u,s) =& g^2 \bar{u}(k_1) P_R u(p_2)\dfrac{1}{u - M_\phi^2} \bar{u}(k_2) P_L u(p_1)\\
\mathcal{M}_s(u,s) =& g^2 \bar{u}(k_1) P_R v(k_2)\dfrac{1}{s - M_\phi^2} \bar{v}(p_2) P_L u(p_1)\,.
\end{align}

The corresponding spin-averaged amplitude squared reads
\begin{align}
    \overline{\abs{\mathcal{M}(u,s)}^2} = &\dfrac{g^4 }{4}\left( \dfrac{\left(u - m_{\rm DM}^2\right)^2}{ \left(u - M_\phi^2\right)^2} + \dfrac{\left(s - m_{\rm DM}^2\right)^2}{ \left(s - M_\phi^2\right)^2}\right)
\end{align}
and the total cross section
\begin{widetext}
\begin{align}
    \sigma_{\rm DM-\nu} =   &\frac{g^4}{64\, \pi\, m_{\text{DM}}^4\, y^2}  \Bigg[2 ( M_\Phi^2 - m_{\text{DM}}^2)\log\left(\dfrac{M_\Phi^2 - \frac{m_{\text{DM}}^2}{y + 1}}{m_{\text{DM}}^2 (y - 1) + M_\Phi^2} \right) \nonumber \\
    &-\frac{(m_{\text{DM}}^2 - M_\phi^2)^2}{m_{\text{DM}}^2 (y - 1) + M_\phi^2}
    - \frac{(y + 1)\, (m_{\text{DM}}^2 - M_\phi^2)^2}{m_{\text{DM}}^2 - M_\phi^2 (y + 1)} + \frac{m_{\text{DM}}^2\, y^2}{y + 1} \nonumber \\
    &+ \frac{m_{\text{DM}}^6\, y^4}{(y + 1)\left(M_\phi^2 - m_{\text{DM}}^2 (y + 1)\right)^2}\Bigg]\,,
\end{align}
\end{widetext}
which in the high-energy limit reads
\begin{align}
    \sigma_{\rm DM-\nu} &\approx \frac{g^4}{32\, \pi\, m_{\text{DM}}^2\, y} 
    + \mathcal{O}\left(\frac{1}{y^2}\right) \,,
\end{align}
supressed by $y^{-1}\propto E_\nu^{-1}$ for $y\gg (m_{Z^{\prime}}/m_{\rm DM})^{2}$, i.e., for high energies.
If we go to the low-energy limit, $y\ll (m_{Z^{\prime}}/m_{\rm DM})^{2}$, we get
\begin{align}
   \sigma_{\rm DM-\nu} \approx \, &\frac{g^4}{32\, \pi\,}\dfrac{y^2\,m_{\text{DM}}^2}{\left(m_{\text{DM}}^2-m_{\Phi}^2\right)^2} 
+ \mathcal{O}\left( y^3 \right)\,,
\end{align}
which grows as $y^2\propto E_\nu^2$.

\subsection{Case 3.1: Real Scalar DM with Fermion Mediator}

The interaction Lagrangian is
\begin{equation}
\mathcal{L} = -g \overline{\chi_R}\, \phi\, \nu_L + \text{h.c.}
\end{equation}
and the scattering amplitude of interest is
\begin{align}
&\mathcal{M}(u,s) = \mathcal{M}_u(u,s) + \mathcal{M}_s(u,s) \nonumber\\
&\mathcal{M}_u(u,s) = g^2 \bar{u}(k_1) P_R \dfrac{\left(\slashed{p}_1 - \slashed{k}_1 + m_{\rm DM}\right)}{u - M_\chi^2} P_L u(p_1) \\
&\mathcal{M}_s(u,s) = g^2 \bar{u}(k_1) P_R\dfrac{\left(\slashed{p}_1 + \slashed{p}_2 + m_{\rm DM}\right)}{s - M_\chi^2} P_L u(p_1) \,.
\end{align}

The spin-averaged amplitude squared reads
\begin{align}
    \overline{\abs{\mathcal{M}(u,s)}^2} = &\dfrac{g^4 \left(s - u\right) \left(m_{\rm DM}^4 - su\right)^2}{2 \left(s - M_\chi^2\right)^2\left(u - M_\chi^2\right)^2} 
\end{align}
and so the total cross section is given by
\begin{widetext}
\begin{align}
    \sigma_{\rm DM-\nu} = &\frac{g^4}{32\, \pi\, m_{\text{DM}}^4\, y^2 \left(M_\chi^2 - m_{\text{DM}}^2 (y + 1)\right)^2} \nonumber \\
    & \Bigg[ m_{\text{DM}}^2 \left(m_{\text{DM}}^2 (y + 1) - M_\chi^2 \right)\left(m_{\text{DM}}^2 \left(y\,(y + 2) + 3\right) - 3 M_\chi^2 (y + 1)\right)   \log\left(\dfrac{m_{\text{DM}}^2 (y - 1) + M_\chi^2}{M_\chi^2 - \frac{m_{\text{DM}}^2}{y + 1}}\right)  \nonumber \\
    & + \frac{m_{\text{DM}}^4 y^2\Big( m_{\text{DM}}^4 \left(y\, (3y\,(y - 1) - 10) - 6\right) + m_{\text{DM}}^2 M_\chi^2 (y + 2)(5y + 6)
    - 6 M_\chi^4 (y + 1) \Big)}{2(y + 1)\left(m_{\text{DM}}^2 (y - 1) + M_\chi^2\right)} \Bigg]\,.
\end{align}
\end{widetext}

In the high-energy limit, the scattering cross section is
\begin{align}
    \sigma_{\rm DM-\nu} &\approx \frac{g^4}{32\, \pi\, m_{\text{DM}}^2\, y} 
    \left( \log\left( \frac{m_{\text{DM}}^2 y}{M_\chi^2} \right) - \frac{3}{2} \right)
    + \mathcal{O}\left(\frac{1}{y^2}\right) \,,
\end{align}
supressed by $y^{-1}\propto E_\nu^{-1}$ for $y\gg (m_{Z^{\prime}}/m_{\rm DM})^{2}$, i.e., for high energies.
If we go to the low-energy limit, $y\ll (m_{Z^{\prime}}/m_{\rm DM})^{2}$, we get
\begin{align}
   \sigma_{\rm DM-\nu} \approx \, &\dfrac{g^4}{16\, \pi}\dfrac{ m_{\text{DM}}^2\, y^4}{\big(m_{\text{DM}}^2 - M_\chi^2\big)^4} + \mathcal{O}\left( y^5 \right)\ ,
\end{align}
which grows as $y^4\propto E_\nu^4$.

\subsection{Case 3.2: Complex Scalar DM with Fermion Mediator}

For the interaction Lagrangian
\begin{equation}
\mathcal{L} = -g \overline{\chi_R}\, \phi\, \nu_L + \text{h.c.}
\end{equation}
the scattering amplitude of interest is
\begin{align}
\mathcal{M}_u(u,s) = g^2 \bar{u}(k_1) P_R \dfrac{\left(\slashed{p}_1 - \slashed{k}_1 + m_{\rm DM}\right)}{u - M_\chi^2} P_L u(p_1) \,.
\end{align}

The spin-averaged amplitude squared is given by
\begin{align}
    \overline{\abs{\mathcal{M}(u,s)}^2} = &\dfrac{g^4 \left(m_{\rm DM}^4 - su\right)^2}{\left(u - M_\chi^2\right)^2}\,. 
\end{align}
Then, the total cross section is
\begin{align}
    \sigma_{\rm DM-\nu} = &\frac{g^4}{32\, \pi }  \Bigg[ \dfrac{y+1}{m_{\text{DM}}^2\, y^2} \log\left(\dfrac{m_{\text{DM}}^2 (y - 1) + M_\chi^2}{M_\chi^2 - \frac{m_{\text{DM}}^2}{y + 1}}\right) \nonumber\\&- \dfrac{1}{m_{\text{DM}}^2 (y - 1) + M_\chi^2}\Bigg]\,,
\end{align} 
which in the high-energy limit is
\begin{align}
    \sigma_{\rm DM-\nu} &\approx \frac{g^4}{32\, \pi\, m_{\text{DM}}^2\, y} 
    \left( \log\left( \frac{m_{\text{DM}}^2 y}{M_\chi^2} \right) - 1 \right)
    + \mathcal{O}\left(\frac{1}{y^2}\right)\,,
\end{align}
supressed by $y^{-1}\propto E_\nu^{-1}$ for $y\gg (m_{Z^{\prime}}/m_{\rm DM})^{2}$, i.e., for high energies.
If we go to the low-energy limit, $y\ll (m_{Z^{\prime}}/m_{\rm DM})^{2}$, we get
\begin{align}
   \sigma_{\rm DM-\nu} \approx \, &\frac{g^4}{64\,\pi}\dfrac{y^2\, m_{\text{DM}}^2}{\,\big(m_{\text{DM}}^2 - M_\chi^2\big)^2} + \mathcal{O}\left( y^3 \right)\ ,
\end{align}
which grows as $y^2\propto E_\nu^2$.

\subsection{Case 4: Scalar DM with Vector Mediator}
Given the interaction Lagrangian
\begin{equation}
\mathcal{L} =\ -g_{\rm DM} \left(\left(\partial^\mu \phi\right)\phi^\dagger - \left(\partial^\mu \phi\right)^\dagger\phi\right) Z^\prime_{\mu} - g_\nu \overline{\nu_L} \gamma^\mu Z^\prime_\mu \nu_L  + \text{h.c.}\,,
\end{equation}
the scattering amplitude is
\begin{widetext}
\begin{align}
&\mathcal{M}_t(t,s) = g_{\rm DM}\, g_\nu\, \bar{u}(k_1) \gamma^\mu P_L\, u(p_1)  \frac{g_{\mu\nu}-\frac{(p_1 - k_1)_\mu(p_1 - k_1)_\nu}{M_{Z^\prime}^2}}{t - M_{Z^\prime}^2} \left(k_2+p_2\right)^\nu
\end{align}
\end{widetext}

Then, the spin-averaged amplitude squared reads
\begin{align}
    \overline{\abs{\mathcal{M}(t,s)}^2} =\dfrac{2\, g_{\rm DM}^2 g_\nu^2 \Big(\left(m_{\rm DM}^2 - s \right)^2+ s\,t \Big)}{\Big(t - M_{Z^\prime}^2\Big)^2}
\end{align}
and the total cros-section is 
\begin{align}
    \sigma_{\rm DM-\nu} = &\frac{g_{\rm DM}^2 g_\nu^2}{8\, \pi\, M_{Z^\prime}^2}\nonumber \\ &\Bigg[1 - \dfrac{M_{Z^\prime}^2\left(y+1\right)}{m_{\text{DM}}^2\,y^2}\log\Bigg(1+\dfrac{m_{\text{DM}}^2y^2}{M_{Z^\prime}^2\left(y+1\right)}\Bigg)\Bigg] \,.
\end{align}

In the high-energy limit, the scattering cross section is
\begin{align}
    \sigma_{\rm DM-\nu} \approx &\frac{g_{\rm DM}^2 g_\nu^2}{8\, \pi}\Bigg[\dfrac{1}{M_{Z^\prime}^2} \nonumber \\ & + \dfrac{1}{m_{\text{DM}}^2\, y}\log\Bigg(\dfrac{M_{Z^\prime}^2}{m_{\text{DM}}^2\, y}\Bigg)\Bigg]    + \mathcal{O}\left(\frac{1}{y^2}\right)\,,
\end{align}
and is constant with energy for $y\gg (m_{Z^{\prime}}/m_{\rm DM})^{2}$, i.e., for high energies.
If we go to the low-energy limit, $y\ll (m_{Z^{\prime}}/m_{\rm DM})^{2}$, we get
\begin{align}
   \sigma_{\rm DM-\nu} \approx \, &\frac{g_{\text{DM}}^2\, g_\nu^2}{16\, \pi}\dfrac{y^2\,m_{\text{DM}}^2}{M_{Z'}^4} + \mathcal{O}\left( y^3 \right)\ ,
\end{align}
which grows as $y^2\propto E_\nu^2$. The bounds for this model are presented in the right panel of Fig.~\ref{fig:bounds_Majoranavector}. We see that attenuation bounds are very similar to Dirac DM, see Fig. 3 of the main text, but DM annihilation limits get much weaker due to the \textit{p}-wave nature of the process.

\subsection{Case 5: Vector DM with Vector Mediator}

For the interaction Lagrangian
\begin{equation}
\mathcal{L} =\ -\frac{1}{2}g_{\rm DM} V^\nu \partial_\nu V^\mu Z^\prime_{\mu} - g_\nu \overline{\nu_L} \gamma^\mu Z^\prime_\mu \nu_L  + \text{h.c.}\,,
\end{equation}
the scattering amplitude is
\begin{widetext}
\begin{align}
&\mathcal{M}_t(t,s) = g_{\rm DM}\, g_\nu\, \bar{u}(k_1) \gamma^\mu P_L\, u(p_1)  \frac{g_{\mu\nu}-\frac{(p_1 - k_1)_\mu(p_1 - k_1)_\nu}{M_{Z^\prime}^2}}{t - M_{Z^\prime}^2} \varepsilon^\nu(p_2)\,p_{2}^\alpha\,\varepsilon^*_\alpha(k_2)\,.
\end{align}

Then, the spin-averaged amplitude squared reads
\begin{align}
    \overline{\abs{\mathcal{M}(t,s)}^2} =\dfrac{ g_{\rm DM}^2 g_\nu^2\, t\,\Big(t - 4\, m_{\rm DM}^2\Big) \Big( m_{\rm DM}^4 - m_{\rm DM}^2 (s + t) + s\left(s + t\right) \Big)}{96\, m_{\rm DM}^4\Big(t - M_{Z^\prime}^2\Big)^2}
\end{align}
and the total cross section is 
\begin{align}
    \sigma_{\rm DM-\nu} = &\frac{g_{\text{DM}}^2\, g_\nu^2}{1536\, \pi\, m_{\text{DM}}^8\, y^2} \Bigg[ \dfrac{m_{\text{DM}}^4 y^2}{2 (y + 1)^2 (m_{\text{DM}}^2 y^2 + M_{Z'}^2 (y + 1))} \nonumber \\
 & \times \Big( m_{\text{DM}}^4 y^3 \left( (y - 13)\, y - 16 \right)
+ m_{\text{DM}}^2 M_{Z'}^2 (y + 1) \left( (7y - 15)\, y^2 + 16 \right)
+ 6 M_{Z'}^4 (y - 1) (y + 1)^2 \Big) \nonumber\\
& + m_{\text{DM}}^2 \left( 4\, m_{\text{DM}}^4 y^2 - 2\, m_{\text{DM}}^2 M_{Z'}^2 (y - 2)^2 - 3 M_{Z'}^4 (y - 1) \right)
 \log\bigg( \frac{m_{\text{DM}}^2 y^2}{M_{Z'}^2\,(y + 1)} + 1 \bigg)  
\Bigg]\,.
\end{align}
\end{widetext}
In the high-energy limit, the cross section is given by
\begin{align}
    \sigma_{\rm DM-\nu} &\approx \frac{g_{\text{DM}}^2\, g_\nu^2}{3072\, \pi\, m_{\text{DM}}^2}
\Bigg[y -
\bigg(8  - 4 \dfrac{M_{Z'}^2}{m_{\text{DM}}^2}\bigg) \log\Bigg( \frac{M_{Z'}^2}{m_{\text{DM}}^2 y} \Bigg)
\nonumber \\ &+ 6 \dfrac{M_{Z'}^2}{m_{\text{DM}}^2} - 15
\Bigg] + \mathcal{O}\left(\frac{1}{y}\right)\,,
\end{align}
grows with $y\propto E_\nu$ for $y\gg (m_{Z^{\prime}}/m_{\rm DM})^{2}$, i.e., for high energies.
If we go to the low-energy limit, $y\ll (m_{Z^{\prime}}/m_{\rm DM})^{2}$, we get
\begin{align}
   \sigma_{\rm DM-\nu} \approx \, &\frac{5\,g_{\text{DM}}^2\, g_\nu^2}{2304\, \pi}\dfrac{y^4\, m_{\text{DM}}^2}{M_{Z'}^4\,} + \mathcal{O}\left( y^5 \right)\ ,
\end{align}
\bibliography{References}

\end{document}